\documentclass[twocolumn,showpacs,preprintnumbers,amsmath,amssymb]{revtex4}
\usepackage{graphicx}
\usepackage{dcolumn}
\usepackage{bm}

\begin{document}
\def\etal{{\it et al.}}
\def\go{\rightarrow  }
\def\be{\begin{equation}}
\def\ee{\end{equation}}
\def\br{\begin{eqnarray}}
\def\er{\end{eqnarray}}
\def\brn{\begin{eqnarray*}}
\def\ern{\end{eqnarray*}}
\def\rf#1{{(\ref{#1})}}
\def\a {{\alpha}}
\def\b {{\beta}}
\def\e {{\epsilon}}
\def\k {{\kappa}}
\def\s {{\sigma}}
\def\w {{\omega}}
\def\sss{\scriptscriptstyle}
\def\nn{\nonumber}
\def\ie{{\em i.e., }}
\def\x{\times}
\def\F {{{\cal F}}}
\def\M {{{\cal M}}}
\def\T {{{\cal T}}}
\def\pb {{\bf p}}
\def\qb {{\bf q}}
\def\kb {{\bf k}}
\def\ket#1{|#1 \rangle}
\def\bra#1{\langle #1|}
\def\Ket#1{||#1 \rangle}
\def\Bra#1{\langle #1||}
\def\up{u_{{\rm p}}}
\def\vp{v_{{\rm p}}}
\def\un{u_{{\rm n}}}
\def\vn{v_{{\rm n}}}

\preprint{}

\title{Neutrino/antineutrino-$^{12}$C charged cross sections in the
projected QRPA formalism}

\author{A. R. Samana, C. A. Bertulani }
\affiliation{%
Department of Physics, Texas A\&M University Commerce,
P.O.3011 Commerce, 75429 TX, USA
}%
\author{and F. Krmpoti\'c$^{1,2,3}$}
\affiliation{$^1$Instituto de F\'{\i}sica La Plata,
CONICET, 1900 La Plata, Argentina}
\affiliation{$^2$Facultad de Ciencias Astron\'omicas y Geof\'{\i}sicas,
Universidad Nacional de La Plata, 1900 La Plata, Argentina,}
\affiliation{$^3$Departamento de F\'{\i}sica, Universidad Nacional de
La Plata, C. C. 67, 1900 La Plata, Argentina}

\date{\today}

\begin{abstract}
The $\nu/\bar{\nu}-^{12}$C cross sections  are evaluated
in the projected quasiparticle random phase approximation (PQRPA).
The cross section for $\nu_e$ as a function of the incident neutrino energy is
compared with recent theoretical calculations of more sophisticated models.
The $\bar{\nu}-^{12}$C cross section is calculated for the
first time with the PQRPA. The distribution of cross sections averaged with
the Michel spectrum as well as with
other estimated fluxes for future experiments
is compared for both $\nu_e$ and $\bar{\nu}_e$. Some astrophysical
implications are addressed.
\end{abstract}


\maketitle

\section{Introduction}

Among the different semileptonic weak interaction with nuclei,
such as charged lepton capture and $\beta^{\pm}$-decays,
the neutrino (antineutrino) scattering is one of most promising
tools for studies of physics beyond the standard model. The massiveness
of neutrinos and the related oscillation are strongly sustained
by many experiment works involving atmospheric, solar, reactor and
accelerator neutrinos ~\cite{Ath96,Ath98,Agu01,Fuk98,Aha05,Ara04,Ahn03}.
Processes such as $\beta$-decay, electron capture and the
double beta decay with two neutrinos are employed to
constraint predictions on  neutrinoless double beta
decay ~\cite{Fas07}. Because the neutrinos interact so weakly
with matter, they are used as messengers from stars and give us useful
information on the possible dynamics of supernova collapse
and explosion as well as on the synthesis of heavy nuclei~\cite{Mcl95,Bal06,Qia07}.
On the other hand, Lazauskas \etal~ have shown in Ref.~\cite{Laz07} that
neutrino-nucleus cross interactions can explore the possibility of
performing nuclear structure studies using low-energy neutrino beams.

The neutrino-nucleus scattering formalism was developed in several references.
For example, the pioneer work of O'Connell \etal~\cite{Con72,Don79}
describes all semileptonic processes, whereas
Kuramotos's formalism \cite{Kur90} only explains the neutrino-nucleus
cross section and additional framework is necessary for the muon
capture rates~Ref.\cite{Luy63}. Krmp\'otic \etal~ have shown in
Ref.~\cite{Krm05} that all these formalisms are equivalent
because they can be described with the same nuclear matrix elements
derived from an effective hamiltonian obtained by carrying out the
Foldy-Wouthuysen transformation and retaining terms up to order
${\cal O}(\kb/M)$, where $\kb$ is the momentum transfer and $M$ is the
nucleon mass.

The neutrino-nucleus scattering on $^{12}$C
is important because this nucleus is a component in many
liquid scintillator detectors. Experiments such as KARMEN~\cite{Mas98,Arm02},
LAMPF~\cite{All90,Kra92} and LSND~\cite{Ath96,Ath98} have used
$^{12}$C to search for neutrino oscillations and to measure
neutrino-nucleus cross sections. As the $^{12}$C nucleus
forms one  of the onion-like shells of a large star before
collapse, it is also important for astrophysics studies. Future
experiments are planning to use $^{12}$C as liquid scintillator, such
as in the spallation neutron source (SNS) at Oak Ridge National
Laboratory (ORNL)~\cite{Efr05}, or in the LVD (Large Volume Detector)
experiment~\cite{Aga07}, developed in the INFN Gran Sasso in Italy.

\begin{figure}[t]
\vspace{-1.5cm}
\begin{center}
{\includegraphics[width=8cm,height=12cm]{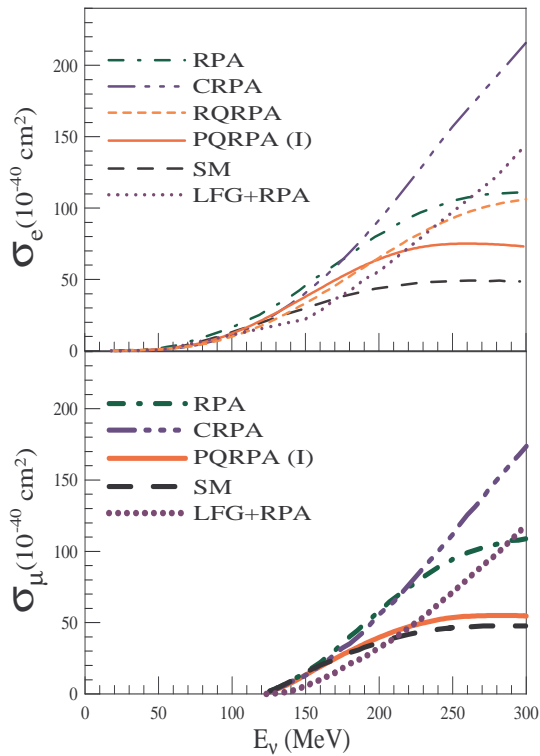}}
\end{center}
\vspace{-1.cm}
\caption{\label{fig1}(Color online) Comparison of the $\sigma_{e,\mu}(E_\nu)$
(top panel) and $\sigma_{\mu}(E_\nu)$ (bottom panel)
as function of $\nu_e$ and $\nu_{\mu})$ neutrino
incident energy for different nuclear structure models
CRPA~\cite{Kol99b}(dashed-dot-dot),
PQRPA~\cite{Krm05}(solid), RPA~\cite{Vol00} (dashed-dot),
SM~\cite{Vol00} (long dashed), LFG+RPA~\cite{Nie04} (dot)
and RQRPA~\cite{Paa07} (short dashed).
}
\end{figure}

There have been great efforts on nuclear structure models to describe
consistently  semileptonic weak processes with $^{12}$C such
as RPA-like models: RPA~\cite{Aue97,Sin98,Vol00},
CRPA~\cite{Kol94,Kol99b,Jac99}, QRPA\cite{Vol00,Sam06},
PQRPA\cite{Krm05,Sam06}, relativistic QRPA
(RQRPA)~\cite{Paa07}; Local Fermi Gas (LFG) plus RPA~\cite{Nie04,Val06},
phenomenological models~\cite{Fuk88,Min82,Min94,Ath06} and the well
known SM~\cite{Vol00,Hay00,Aue02}.
Figure \ref{fig1} summarizes the state of the art of the most recent
electron and muon neutrinos cross section on $^{12}$C as function of
the neutrino energy for several nuclear structure calculations.
The residual interaction used in those calculations is not unique and
it varies from the simple $\delta$-interaction in PQRPA~\cite{Krm05},
Skyrme-type effective interaction in QRPA- SM- RPA~\cite{Vol00},
G matrix for the Boon (or Landau Migdal) potential in CRPA~\cite{Kol99b},
meson-exchange density dependent relativistic mean field effective
interactions - DD-ME2 and finite range Gogny interaction in RQRPA~\cite{Paa07},
and nucleon-nucleon effective force supplemented by nucleon-$\Delta$(1232) and
$\Delta$(1232)-$\Delta$(1232) interactions in LFG+RPA~\cite{Nie04}.
From Figure~\ref{fig1} we note that the behavior of $\sigma$ from both
$\nu_e$-$^{12}$C and $\nu_\mu$-$^{12}$C reactions evaluated in SM, RPA and PQRPA
diverges substantially from CRPA and LFG+RPA with increasing neutrino energy.
One sees that, while within the SM and the PQRPA both
$\sigma_\ell$ ($\ell= e, \mu$) start to level at around 200 MeV, the RPA
does the same but around 300 MeV. The CRPA  and LFG+RPA
are those models where the cross sections continue to increase at these energies.
It should be pointed out, however, that all three RPA-like
calculations used the same  single-particle space  and
only 1p-1h (2 quasiparticles) excitations have been considered.
The major differences comes from the type of correlations included in
each case. In Ref.~\cite{Krm05} it was noted that the similarity between the SM
and the PQRPA results, as well as  the difference with the RPA
calculations, can be attributed to the Pauli Principle. But, it
is hard to understand the qualitative difference between the RPA and CRPA.
It could happen that some additional effects, such as contributions of
high-lying single-particle states or of 2p-2h (4 quasiparticles)
excitations, become important for neutrino energies above the DIF energy
region, preventing in this way the leveling of  $\sigma (E_\nu)$. This is
an important issue and  worth to be analyzed, in particular in view
of the recent LFG+RPA results in the QE (quasi-elastic) region
at intermediate energies, where $\sigma$ becomes flat only at about 1 GeV.
This is an open question that should be considered in future studies.

Refs.~\cite{Krm02,Krm05,Sam06} have shown that in order to describe the weak decay
observables  in  a light $N=Z$ nucleus as such as $^{12}$C in the framework
of the RPA model one must, besides including the BCS correlations,
also include the particle number projection procedure.
Recently in Ref.~\cite{Sam08}, the PQRPA was used
to calculate the $^{56}$Fe$(\nu_e,e^-)^{56}$Co cross section. The resulting
cross section was compared with a QRPA calculation with the same
interaction showing that the projection procedure is important for
medium mass nuclei. In heavy nuclei, where
the neutron excess is usually large,
the projection procedure is less important~\cite{Krm93}.

In this work the neutrino and antineutrino-nucleus $\sigma (E_\nu)$
cross section are evaluated with the formalism developed in
Ref.~\cite{Krm05} for neutrino-nucleus reactions using the PQRPA
as the nuclear structure model. We adopt that formalism because
it is best suited for a nuclear structure comparison.
With a minor modification on this formalism in the
leptonic traces, we evaluate the antineutrino reactions.
As the PQRPA model solves the inconveniences that appear
in applying  RPA-like models to describe the nuclear structure of the
$\{^{12}{\rm B},^{12}{\rm C},^{12}{\rm N}\}$ triad,
we calculate the $^{12}$C$({\bar \nu}_e,e^+)^{12}$B cross section
with the nuclear matrix elements (NME) in this model.

It was shown in \cite{Sam06} that large multipoles
are important in the detection window  $E_e \in [60-200]$ MeV
of LSND for $\nu_\mu \go \nu_e$ neutrino oscillations.
This yields an enhancement of the
$\nu_\mu \go \nu_e$ oscillation probability.
Recent results of $\sigma (E_\nu)$ for different multipolarities
with RQRPA~\cite{Paa07} allow us to compare with those provided
by PQRPA. For the sake of completeness the partial contribution
to the inclusive cross section for ${\bar \nu}_e-^{12}$C  is analyzed.
One alternative method to study the large
multipoles was discussed by Lazauskus and Volpe in Ref.~\cite{Laz07}
through the so called low-energy neutrino beta-beams. Different distributions
of the averaged cross sections on nuclei as $^{16}$O, $^{16}$Fe, $^{100}$Mo,
and $^{208}$Pb were presented and the feasibility of using beta-beams
was clarified. With that work we have learned that low-energy neutrino beams
could provide information on the forbidden states, in particular the
spin-dipole. The QRPA model was employed to describe
nuclei beyond the closed-shell approximation. However,
QRPA predictions of Ref.~\cite{Vol00} do not yield good results for
$^{12}$C because the configuration mixing is not properly accounted
for and the projection procedure (as done in Ref.~\cite{Krm05}) is not
included.
We have mentioned previously that $^{12}$C
is reasonably well described with the PQRPA and so, we can expand the
discussion of the $\beta$-beam to this nucleus.
In spite of this fact we fold the cross section with the DAR flux
and those from beta-beams and compare these results.
Some topics on astrophysical applications with the
resulting cross sections are addressed.

In Sec. II we briefly overview the formalism for the neutrino-nucleus
cross sections as well the PQRPA.  In Sec. III we compare the
$\nu/\bar{\nu}-^{12}$C cross sections with other RPA-like models.
Summarizing conclusion are presented in Sec. IV.

\section{Formalism Neutrino-Nucleus}

The cross section
for $\nu_e + (Z, A)\go (Z+1, A)+ e^-$, as a function of the
incident neutrino energy for each nuclear spin, is given by
\br
\s(E_e,J_f)& = &\frac{|\pb_e| E_e}{2\pi} F(Z+1,E_e)
\int_{-1}^1
d(\cos\theta)\T_{\s}(|\kb|,J_f),\nn\\
\label{1}\er
where $F(Z+1,E_e)$ is the usual scattering Fermi
function, $k=p_e-q_\nu$ is the momentum transfer, $p_e$ and $q_\nu$
are the corresponding electron and neutrino momenta, and
$\theta\equiv \hat{\qb}_\nu\cdot\hat{\pb}_e$ is the angle between
the incident neutrino and emerging electron. The $\sigma(E_e, J_f)$
cross sections are obtained within first-order perturbation theory
according to Ref.~\cite{Krm05}, where velocity-dependent terms are
included in the weak effective Hamiltonian. The transition amplitude
$\T_{\s}(|\kb|,J_f)$ depends on the neutrino leptonic traces and on
the nuclear matrix elements (NME), as explained in
Ref.~\cite{Krm05}. They are evaluated in the PQRPA .

In Refs.~\cite{Krm02,Krm05,Sam08} we evaluated the
neutrino-nucleus reaction  $\nu_e + (Z, A)\go (Z+1, A)+e^-$ for $^{12}$C.
To evaluate the antineutrino-nucleus
reaction  ${\bar \nu}_e + (Z, A)\go (Z-1, A)+ e^+$ with this formalism
it is necessary to modify the lepton trace $\L_{\pm1,\pm1}$, that
appears in \cite[(2.35)]{Krm05}, to
\br
\L_{\pm1,\pm1}&=&1-\frac{q_0p_0}{E_\ell E_\nu}\pm
\left(\frac{q_0}{E_\nu}-\frac{p_0}{E_\ell}\right)S_1,
\label{2}\er
where the factor $S_1=\pm 1$ for neutrino/antineutrino reactions.
The inclusive $(\nu/\bar{\nu})$-nucleus cross section reads
\be
\s_\ell^{\rm inc}(E_{\nu})=\sum_{J^{\pi}_f}
\s_\ell(E_\ell=E_\nu-\w_{J_{f}^\pi},J_{f}^\pi);\:\;
\ell=\left\{
\begin{array}{l}
e,\;\; \mbox{for}\; e^-,
\\
{\overline e},\;\; \mbox{for}\; e^+.
\\\end{array}\right.
\label{3}\ee
The spin and parity dependent cross section $\s_\ell(E_\ell,J_{f}^\pi)$
is given by equation \rf{1} (explicitly by \cite[(2.19)]{Krm05}) with
the additional modification \rf{2}; $\w_{J_{f}}$ are the excitation
energies  for each nuclear state in the daughter nuclei $(A,N\pm1)$
(the '$+$' for neutrino-nucleus reaction
and the '$-$' for antineutrino one) relative to the
ground state in the parent nuclei $(A,N)$.

The flux averaged cross section reads
\be
\overline{\s}_\ell= \int dE_{\nu}
\s_\ell(E_\nu) n_\ell(E_{\nu}),
\label{4}\ee
where $\s_e(E_{\nu})~(\s_{\overline e}(E_{\nu}))$,
is the neutrino~(antineutrino) cross section
as a function of the neutrino~(antineutrino) energy eq.~\rf{3} and
$n_\ell(E_{\nu})$ is the neutrino~(antineutrino) normalized flux.
In Refs.~\cite{Krm02,Krm05} we have folded the $\s_e(E_\nu)$ with the
Michel energy spectrum~\cite{Kol99a,Arm02}
\[
n_e(E_\nu)=\frac{96 E_\nu^2}{M_\mu^4} \left(M_\mu-2 E_\nu\right),
\]
where $M_\mu$ is the muon mass, in . This neutrino flux is
normalized to one in the DAR (decay-at-rest) energy interval.
In this work, we fold the antineutrino
cross section, $\s_{\overline e}(E_{\nu})$, with
antineutrino fluxes from conventional DAR source~\cite{Arm02},
and with those from the decay of $^6$He ions boosted
at $\gamma=6$, $\gamma=10$,
and $\gamma=14$ presented in Fig.~4 of Ref.~\cite{Laz07} .
Specific details on the neutrino fluxes associated to low-energy
$\beta$-beams are given in Refs.~\cite{Laz07,Ser04}.

The formalism of the PQRPA was developed in
Refs.~\cite{Krm93, Krm05}.
When the excited states $\ket{J_f}$
in the final $(Z\pm 1,N\mp 1)$ nuclei
are described within the PQRPA, the transition amplitudes
for the multipole charge-exchange operator ${\sf Y}_{J}$, read
\br
\Bra{J_f, Z+\mu,N-\mu}{\sf Y}_{J}\Ket{0^+}  =  {1 \over
(I^{Z}I^N)^{1/2}}
\nn\\
\sum_{pn}
 \left[ {\Lambda_\mu(pnJ) \over
 (I^{Z-1+\mu}(p)I^{N-1+\mu}(n))^{1/2}}
 X_{\mu}^{\ast}(pnJ_f)\right. +
 \nonumber\\
 +\left.{\Lambda_{-\mu}(pnJ) \over
 (I^{Z-1-\mu}(p)I^{N-1-\mu}(n))^{1/2}} Y_\mu^{\ast}(pnJ_f)\right],
\label{5} \er
with the one-body matrix elements given by
\begin{eqnarray}
&&\Lambda_\mu(pnJ)=-\frac{\Bra{p}{\sf Y}_{J}\Ket{n}}{\sqrt{2J+1}}
 \left\{
\begin{array}{l}
\up \vn,\;\; \mbox{for}\; \mu = +1 \\ \un \vp,\;\; \mbox{for}\; \mu
= -1 \
\\\end{array}\right.,
\label{6}
\end{eqnarray}
where
\br
I^{K}(k_1k_2\cdot\cdot k_n)&=& \frac{1}{2\pi i}
\oint \frac{dz}{z^{K+1}} \s_{k_1}\cdots \s_{k_n}
\nn\\
&&\x \prod_{k}(u_k^2 + z^2 v_k^2)^{j_k+1/2};
\nn\\
\s_{k}^{-1}&=&u^2_k+z^2_kv_k^2,
\label{7}\end{eqnarray}
are the  PBCS number projection integrals, and
$(u_k, v_k,)$ are the usual occupation amplitudes
of the $k$-level.
The forward, $X_{\mu}$, and backward, $Y_{\mu}$,  PQRPA amplitudes
are obtained by solving the RPA equations, as explained in Ref.~\cite{Krm05}.
It is possible to recover the usual QRPA from the PQRPA
dropping the index $\mu$ in the RPA matrixes and taking
the limit $I^K \go 1$, and substituting the unperturbed  PBCS energies
by the BCS energies relative to the Fermi level.
It is  also necessary to impose the subsidiary conditions
$Z = \sum_{j_p} (2j_p+1)^2 v_{j_p}^2$ and
$N = \sum_{j_n} (2j_n+1)^2 v_{j_n}^2$
to average the number of particles because they
are no longer good quantum numbers.

\section{Numerical results and discussion}

In this section, our theoretical results for the cross
section $\nu_e + ^{12}C \rightarrow {^{12}N + e^-}$
and ${\bar \nu}_e + ^{12}C \rightarrow {^{12}B + e^+}$
within the PQRPA are compared with other RPA-type model.
As with our previous work~\cite{Sam08},
we employ the $\delta$-interaction (in MeV fm$^3$)
\[
V=-4 \pi \left(v_sP_s+v_tP_t\right) \delta(r),
\]
with  different coupling constants $v_s$ and $v_t$ for the
particle-hole, particle-particle, and pairing channels.
This interaction leads to a good description of single and
double $\beta$-decays and it has been used extensively
in the literature~\cite{Hir90a,Hir90b,Krm92,Krm94}.
The configuration space includes
the single-particle orbitals with $nl=(1s,1p,1d,2s,1f,2p)$
for both protons and neutrons. The s.p. energies, pairing
strengths and projection procedure are detailed in
Tables III and VI of Ref.~\cite{Krm05}. The single-particle wave
functions were also approximated with those of the HO with the
length parameter $b=1.67$fm, which corresponds
to $ \hbar \omega=45A^{-1/3}-25A^{-2/3}$~ MeV for the oscillator energy.
In our previous works Refs.~\cite{Krm02,Krm05} we have also pointed out
that the values of the coupling strengths $v_s$ and $v_t$ within the
$pp$ and  $ph$ channels used in $N > Z$ nuclei
($v^{pp}_s\equiv v^{pair}_s$, and
$v^{pp}_t\gtrsim v^{pp}_s$),  might not be suitable for
 $N=Z$ nuclei. Then, the best agreement with data in $^{12}$C
(energy of the ground state in $^{12}$C,
 $B(GT)$ of $^{12}$N($\beta^+$)$^{12}$C, and exclusive muon capture on
$^{12}$B $\equiv \lambda^{exc}(1^+_1)$) is obtained when the pp
channel is totally switched  off, \ie $v^{pp}_s\equiv v^{pp}_t=0$.
Three different set of values for the $ph$ coupling
strengths with physical meaning are~\cite{Krm02}:
P(I): $v^{ph}_s=v^{pair}_s=24$ MeV~ fm$^3$ and
$v^{ph}_t=v^{ph}_s/0.6=39.86$ MeV~ fm$^3$;
P (II): $v^{ph}_s=27$  MeV~ fm$^3$ and  $v^{ph}_t=64$ MeV~ fm$^3$; and
P (III): $v^{ph}_s=v^{ph}_t=45$ MeV~ fm$^3$.

Among the different models that studied the observables in the
$\{^{12}{\rm B},^{12}{\rm C},^{12}{\rm N}\}$ triad, some countable
examples give an estimate of the systematic error on the cross section
as a whole, \ie based on the measured observables in the triad. The
LSND experiment~\cite{Ath98} has used
the CRPA cross section with systematic uncertainties of
($+22\%,-45\%$) folded with the muon
neutrino fluxes. When the PQRPA was used to reanalyze the LSND data
for the $\nu_\mu \go \nu_e$ oscillation search, an uncertainty on
the folded cross section was $\pm28 \%$, where only $\approx 20\%$ was
based on theoretical uncertainties.
Here, we do not pretend to make
a detailed study of the uncertainties of the model such as that
developed by Valverde \etal in Ref.~\cite{Val06} for the LFG+RPA model,
instead we are going to establish a simple criteria of uncertainty
for the parametrization of residual $\delta$-interaction
in the PQRPA using the experimental data available on the literature.
\begin{table}[h]
\caption{Weak observables reproduced in the
PQRPA model for the
$\{^{12}{\rm B},^{12}{\rm C},^{12}{\rm N}\}$ triad in comparison
with the experimental values. In the last three lines, we show
the deviation parameters  $\eta_{\sss A}, \eta_{\sss B}$
and ${\overline \epsilon}$ (in $\%$) described in the text. The $B(GT)$
is the averaged value for $B(GT_+)$ and  $B(GT_-)$. The $\lambda$-values are in
units of $10^3$ s$^{-1}$, ${\overline \s}_e$ and ${\overline \s}_\mu$-values
in units of $10^{-42}$ cm$^2$ and $10^{-40}$ cm$^2$ respectively, $E_{gs}(^{12}N)$
is in MeV and $B(GT)$ is dimensionless.
}%
\label{tab:1}
\newcommand{\cc}[1]{\multicolumn{1}{c}{#1}}
\renewcommand{\tabcolsep}{0.2pc} 
\renewcommand{\arraystretch}{1.2} 
\bigskip
\begin{tabular}{c|ccc|c}\hline
Obs.                 &  P(I) &P(II) &P(III)&Exp           \\ \hline
$\lambda^{exc}(1^+_1)$& 7.52  &6.27  & 6.27 &$6.2 \pm   0.3 $ \cite{Mil72} \\
$\lambda(1^-_1)$     & 1.06  &0.49   & 0.98 &$0.62\pm   0.2 $ \cite{Mea01,Sto02}\\
$\lambda(2^-_1)$     & 0.31  &0.18   & 0.16 &$0.18\pm   0.1 $ \cite{Mea01,Sto02}\\
$\lambda^{inc}$      & 48.16 &42.56  &44.67 &$38  \pm   1   $ \cite{Suz87}\\
${\overline \s}_e^{exc}$
                     & 9.94  &8.07  & 8.17 &$8.9  \pm  0.9 $ \cite{Aue01} \\
${\overline \s}_e^{inc}$
                     & 21.67 &18.6  & 17.54&$13.2 \pm   0.7 $ \cite{Aue01}\\
${\overline \s}_{\mu}^{exc}$
                     & 0.74  &0.59  & 0.59 &$0.56 \pm   0.13$ \cite{Aue02a}\\
${\overline \s}_{\mu}^{inc}$
                     & 14.69 &12.94 &13.51 &$10.6 \pm   1.8 $ \cite{Aue02a}\\
$E_{gs}(^{12}N)$     & 17.89 &18.14 &18.13 &$17.3381\pm 0.001$\cite{Ajz85}\\
$B(GT)$              & 0.568 &0.477  &0.48 &$0.496  \pm 0.030$ \cite{Alb78}\\
\hline
$\eta_{\sss A}$             & 175    &253  &250   &\\
$\eta_{\sss B}$             & 5.7    &3.0  &3.1   &\\
${\overline \epsilon}$ (\% )& 35     &12   & 17   &\\
\hline
\end{tabular}
\end{table}
Table \ref{tab:1} shows a summary of the weak observables
as described in the PQRPA model for
$\{^{12}{\rm B},^{12}{\rm C},^{12}{\rm N}\}$ triad
in previous References \cite{Krm02,Krm05}. As a measure of
how good is the parametrization employed in this model,
we define the parameters of deviation from the experimental
values as
\br
\eta&=&\sqrt{\frac{1}{N_0}
\sum_{n=1}^{N_0}\left(\frac{y_{cal}(n)-y_{exp}(n)}{\delta y_{exp}(n)}\right)^2},
\nn \\
{\overline \epsilon} &=&
\left[\sum_{n=1}^{N_0}\left(\frac{y_{cal}(n)-y_{exp}(n)}{y_{exp}(n)}\right)\right]
\frac{100 \%} {N_0}.
\label{8}\er

These parameters were evaluated with the observables of Table~\ref{tab:1},
$y_{cal/exp}(n)\equiv \{\lambda^{exc}(1^+_1), ~...~ ,B(GT)\}$, with $N_0=10$
for $\eta_{\sss A}$, and $N_0=9$ for $\eta_{\sss B}$ where $E_{gs}(^{12}N)$
is excluded. From the $\eta_{\sss A}$ values we note that the P(I)
parametrization achieves the lower value if we conserve the $E_{gs}(^{12}N)$.
For $\eta_{\sss B}$ values of  P(II) and P(III) we noted that they are of
the same order and that the deviation in the ground state energy increases
the partial contribution in the sum due to its lower experimental error.
We can conclude from the values of ${\overline \epsilon}$
that the PQRPA results overestimate in $\approx 21 \%$ the
experimental values.
We remark that the parameters of the residual interaction, $v_T^{PP}$
(in $pp$ channel) and $v_s^{PH}$ (in $ph$ channel), were fixed to reproduce
only the $E_{gs}(^{12}N)$ and $B(GT)$, whereas the other obtained observables
are predictions of our model. In this way, we estimate
and analyze the antineutrino cross section with P(I). Their results are
considered as a upper limit for PQRPA. The P(II) results are
better estimates for this model.

Let us summarize some interesting issues for the exclusive cross
section, $\sigma^{exc}$, that involves only the transitions to the ground
state, in the view of some future experiments that will use
 $^{12}$C as scintillator liquid detector to search signals of
supernovae neutrinos. In the LVD experiment~\cite{Aga07}
it was estimated that the $(\nu_e+{\bar \nu}_e)$ interactions
on $^{12}$C affect the total detected signal. The LVD detector
use exclusive cross
sections, \ie $^{12}{\rm C}_{gs}(O^+)(\nu_e, e^-)^{12}{\rm B}_{gs}(1^+)$
and $^{12}{\rm C}_{gs}(0^+)({\bar \nu}_e, e^+)^{12}{\rm B}_{gs}(1^+)$,
from the EPT (Elementary Particle Treatment) model~\cite{Fuk88} to
compare with the expected signal in the detector.
This theoretical description of exclusive cross section was used for
estimate possible events of supernovae neutrinos in experiments
like Borexino~\cite{Cad02} and LENA~\cite{Und06}.
In all the different variation on the parameters studied by the LVD group,
the cross section was kept fixed within error. It could be reasonable for
the exclusive neutrino cross section $^{12}$C, that is theoretically and
experimentally well determined. The exclusive folded cross section for
$(\nu_e,^{12}C)$ $\overline{\s}_e^{\rm exc}\equiv\overline{\s}_e(J^{\pi}_f=1^{+}_1)$
was measured by LAMPF~\cite{All90,Kra92}, KARMEN Collaboration~\cite{Mas98}
and LSND Collaboration~\cite{Ath96,Ath98} in the DAR region.
In general, theoretical estimates for ${\overline \s}^{\rm exc}_e$
are in agreement with the experimental data. For this reason, this
cross section was used to calibrate different experiments
leaving a faithful knowledge of neutrino fluxes in order to search
neutrino oscillations~\cite{Agu01,Arm02}. It was also employed to
demonstrate the feasibility of observing low-energy neutrino
induced transitions between well determined nuclear states.

\begin{figure}[t]
\begin{center}
{\includegraphics[width=9.cm,height=11.cm]{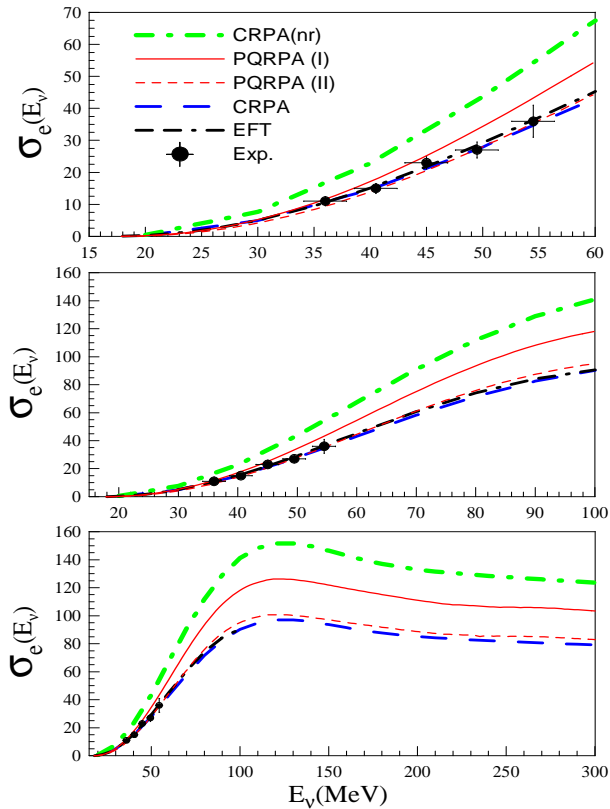}}
\end{center}
\vspace{-1cm}
\caption{\label{fig2}(Color online) Exclusive $\nu_e-^{12}$C
cross section in different
nuclear structure models, CRPA and CRPA(nr)~\cite{Kol94},
PQRPA~\cite{Sam06} and EPT~\cite{Fuk88}. In the top panel
theoretical cross sections are compared with the
experimental data of Ref.~\cite{Aue02} in the DAR region.
The middle and bottom panels show the cross sections
for other energy intervals, as described
in the text.
}%
\end{figure}
Figure \ref{fig2} shows the
$\sigma_e(J_f^\pi=1^+_{gs})$ as function of the incident neutrino
energy for PQRPA \cite{Sam06}, CRPA
\cite{Kol94} and EPT \cite{Fuk88} on the energy interval important for
search of neutrino oscillations (top panel),
for supernovae neutrinos (middle panel), and to
the upper limit of 300 MeV (bottom panel).
CRPA used a reduction factor average of $\approx 1.5$ to
reproduce the $B(GT)$ of $^{12}$N($\beta^+$)$^{12}$C and
$^{12}$B($\beta^-$)$^{12}$C together with the exclusive muon capture
in $^{12}$B $\equiv \lambda^{exc}(1^+_1)$.
This factor ensures that
CRPA cross section is between the experimental error. CRPA~(nr) non-reduced
means that the CRPA cross section is multiplied by $1.5$.
The PQRPA reproduces the
cross sections without need of a reduction factor.
P(I) overestimates the experimental data over 45 MeV
and P(II) is as good as the CRPA reduced and EPT~\cite{Fuk88}.
Nevertheless, there is no need for a
reduction factor with the large multipoles. The
neutrino inclusive cross section with and without reduction factors
does not present differences when the energy increases.
In the middle panel of Figure \ref{fig2} we shown $\sigma(1^+_1)$
in the energy interval that will search for supernovae neutrinos. We note
that the P(II), EPT and CRPA results are close in the DAR region.
From 60 MeV to 100 Mev and up to 300 MeV they separate
progressively with increasing neutrino energy.

\begin{figure}[t]
\vspace{-.5cm}
\begin{center}
{\includegraphics[width=7.5cm,height=11cm]{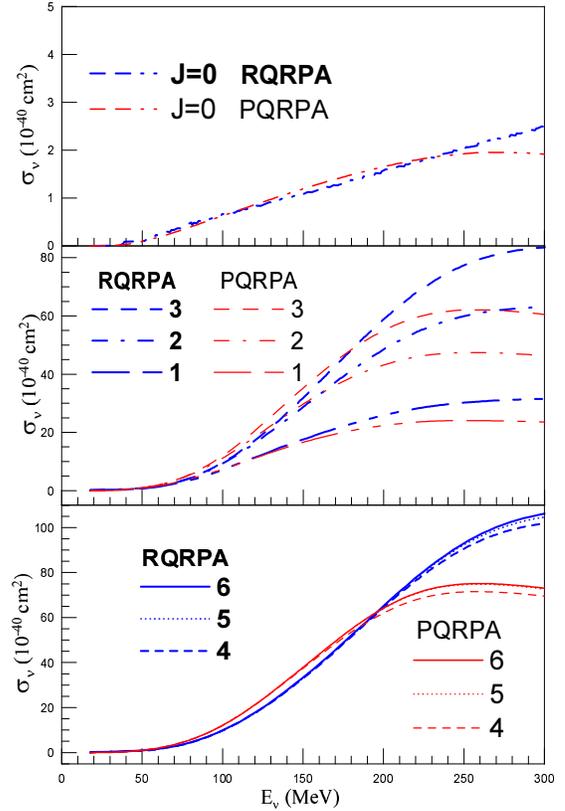}}
\end{center}
\vspace{-.5cm}
\caption{\label{fig3}(Color online) Comparison of the $\sigma_e(E_\nu)$ in
PQRPA and RQRPA~\cite{Paa07} models for
different multipoles showed according the increase nuclear spin from
$J_{\sss min}=0^{\pm}$ to $J_{\sss max}=6^{\pm}$.}
\end{figure}
Figure \ref{fig3} compares the RQRPA~\cite{Paa07} and PQRPA neutrino
cross sections as a function of the neutrino energy
according the increase nuclear spin from
$J_{\sss min}=0^{\pm}$ to $J_{\sss max}=6^{\pm}$, \ie
\br
J=0&\equiv&\sigma_e(0^+)+\sigma_e(0^-),
\nn\\
J=1&\equiv&\sigma_e(0^+)+\sigma_e(0^-)+\sigma_e(1^+)+\sigma_e(1^-),
\nn\\
&\therefore&
\nn\\
J=6&\equiv&\sum_{J^\pi_f=0^+,0^-}^{J^\pi_f=6^+,6^-} \sigma_e(J^\pi_f),
\label{9}\er
in the same way as in Ref.~\cite{Paa07}.
Our PQRPA results are similar to those obtained from the RQRPA.
The largest contribution comes from $J=1^{\pm}$ and $J=2^{\pm}$
and the contribution of higher multipolarities gradually decreases.
In other words they are due to the
allowed $\sigma_e(0^+, 1^+)$, first forbidden
$\sigma_e(0^-, 1^-, 2^-)$, second forbidden  $\sigma_e(2^+ ,3^+)$,
third forbidden  $\sigma_e(3^-, 4^-)$,
fourth forbidden $\sigma_e(4^+, 5^+)$ transitions (hereafter
allowed transitions (AT) and forbidden transitions (FT)), where
the contribution from the higher forbiddenness
is decreasing gradually.
In particular for $J=0$ and $J=1$ both models have
similar cross section up to $\approx 150$ MeV. With increasing
$J$ the differences in these models increase
starting from lower neutrino energies.
From the bottom panel of Fig.~\ref{fig2}
we see that the inclusive cross sections $J=6$ begin
to level out at 200 MeV for PQRPA and 300 MeV for RQRPA.
Comparing the inclusive folded cross section in DAR
region we have $18.6 \x 10^{-42}~{\rm cm}^2$ in PQRPA (II) and
$12.14 \x 10^{-42}~{\rm cm}^2$ for RQRPA. We note that RQRPA is close
to the experimental value $13.2\pm0.7 \x 10^{-42}~{\rm cm}^2$~\cite{Aue01}.
But the situation is inverted for the DIF region for
$(\nu_\mu, \mu^-)$, where the $\overline{\s}_\mu^{\rm inc}$ is
$12.9 \x 10^{-40}~{\rm cm}^2$ in PQRPA (PII) and
$19.59 \x 10^{-40}~{\rm cm}^2$ for RQRPA, \ie the PQRPA is closer to the
experimental value $12.4\pm0.7 \x 10^{-40}~{\rm cm}^2$~\cite{Aue01}.
In tables VI and VII of \cite{Krm05} the cross sections
$\overline{\s}_{e,\mu}(J^\pi_f)$ for each final state
with spin and parity $J^\pi_f$, as well as the exclusive,
$\overline{\s}_{e,\mu}^{\rm exc}\equiv\overline{\s}_{e,\mu}(J^{\pi}_f=1^{+}_1)
$, and inclusive
$\overline{\s}_{e,\mu}^{\rm inc}=\sum_{J^{\pi}_f}\overline{\s}_e(J^{\pi}_f)$
are presented.
It should be remembered that the main contribution
to $\overline{\s}_e^{\rm inc}$ in the DAR region comes  essentially
from the ground state ($\approx 67\%$), whereas in the DIF region
the forbidden transition are most important. Then it could be interesting
to compare the contribution of
different multipoles to the folded cross sections
for PQRPA vs RQRPA in the DAR and DIF region.

\begin{figure}[h]
\begin{center}
{\includegraphics[width=8cm,height=11.cm]{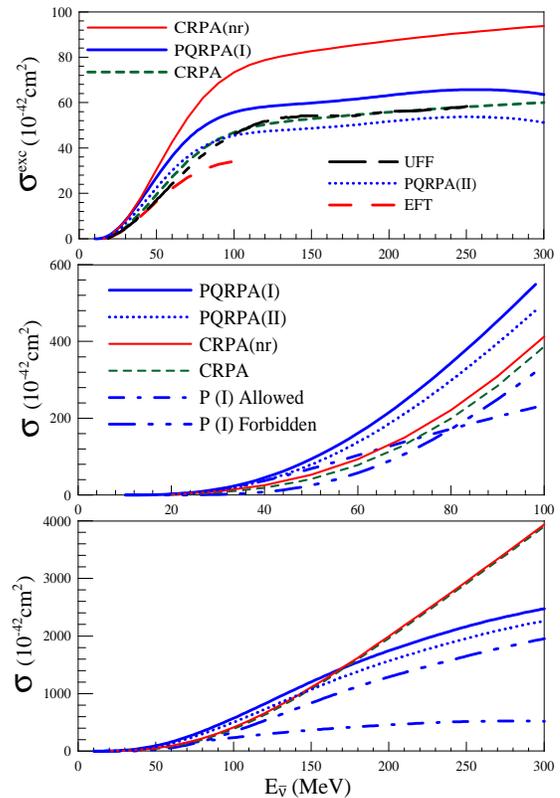}}
\end{center}
\vspace{-1.cm}
\caption{\label{fig4}(Color online)
${\bar \nu}_e-^{12}$C cross sections  as a function
of the incident antineutrino energy with different
nuclear structure models.
(top panel) Exclusive cross sections: PQRPA(I) (solid) and (II) (dotted),
CRPA plus partial occupations (short dashed) and
CRPA(nr) plus partial occupations non-reduced (thin solid)~\cite{Kol99b}
, UFF (short-large dashed)~\cite{Min82}, EPT (large dashed )~\cite{Fuk88}.
In the middle and bottom panels we compare the inclusive cross sections
for PQRPA (I)(solid) and (II) (dotted)
, CRPA+PO (short dashed) and CRPA+PO(nr) (thin solid)~\cite{Kol99b}
for other neutrino energy  interval, as described in the text. For
PQRPA (I), we present the allowed (dashed-dot) and forbidden
(dashed-dot-dot) contributions.
}
\end{figure}
Experimental data on exclusive and other excited
states necessary to build the inclusive cross section for
antineutrino cases are scarce or null.
Figure \ref{fig4} illustrates the
${\bar \nu}_e-^{12}$C cross sections  as a function
of the incident antineutrino energy in different
nuclear structure models: PQRPA (I) (solid) and (II),
CRPA+partial occupations (PO) Ref.~\cite{Kol99b}
and CRPA+PO(nr) non-reduced. 
Also shown are the cross sections from the
uncorrelated factor form (UFF) of Ref.~\cite{Min82}
(it is another EPT calculation) and EPT~\cite{Fuk88}.
In the top panel we compare the antineutrino exclusive cross
sections, $\sigma_{\overline e}^{\rm exc}$, where the PQRPA (I) is larger
that other ones, if the CRPA+PO(nr) is not taken into account.
Below $100$ MeV PQRPA (II) is only slightly larger than the
other models, and it becomes smaller beyond this energy. In particular,
the EFT model is the smallest below 100 MeV, when the other approximations
are very close~(EFT separates of the group above 55 MeV). This
behavior of the EFT antineutrino exclusive cross section of Ref.~\cite{Fuk88}
is different of what is obtained in the neutrino case (see middle panel
of Fig.~\ref{fig2}) and it could have important
consequences for experiments, such as Borexino, LENA  and LVD,
which use this cross section for the detection
of supernovae neutrinos.
The middle panel of Fig.~\ref{fig4} shows $\sigma_{\overline e}$
in the energy interval that could be
relevant for supernovae neutrinos. In this region the
inclusive cross sections show the characteristic
increase with $E_\nu^2$. The allowed and forbidden
contributions for the inclusive cross section are shown for
PQRPA (I). We note that below energies of $\approx 80$ MeV the main
contribution to $\sigma_{\overline e}^{\rm inc}$ comes from the
AT (showing a change of convexity) and above
this value the FT grows faster that the other ones.
In the bottom panel of Fig.~\ref{fig4} the mentioned behavior for the
forbidden cross sections remains the same through 300 MeV, whereas
the allowed cross section begins to saturate at $\approx 200$ MeV.
The forbidden and inclusive PQRPA cross sections still grow slowly
and apparently level out beyond $300$ MeV.
Comparing the $\sigma_{\overline e}^{\rm inc}$
from PQRPA and CRPA(and CRPA(nr)) we note that the PQRPA
results are larger than the CRPA's ones below
the crossing point at $\approx 170$ MeV. Above this energy the
slope of these cross sections is so different,
$\sigma_{\overline e}^{\rm inc}$ of CRPA increases almost
linearly and $\sigma_{\overline e}^{\rm inc}$ of PQRPA increases
slowly, separating from CRPA.
\begin{figure}[t]
\begin{center}
\vspace{-2cm}
{\includegraphics[width=8.cm,height=9.cm]{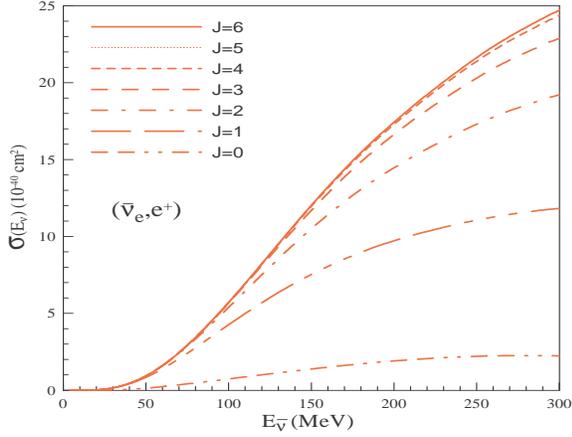}}
\end{center}
\vspace{-1.5cm}
\caption{\label{fig5} ${\bar \nu}_e-^{12}$C cross section, $\sigma(E_{\bar \nu})$,
as a function of the $E_{\bar \nu}$ energy in the
PQRPA (parametrization P(I)) model for
different multipoles showed according to increasing
nuclear spin, from $J_{\sss min}=0^{\pm}$ to $J_{\sss max}=6^{\pm}$.}
\end{figure}
Figure~\ref{fig5} shows the $\s_{\overline e}(E_{\nu})$
with increasing nuclear spin, according to equation~\rf{9}.
As with the neutrino case, the largest contribution comes
from $J=1^{\pm}$ and $J=2^{\pm}$
and the contribution from higher forbiddenness fades out.

\begin{table*}[t]
\caption{\label{tab:2}~Calculated flux-averaged cross section in partial fraction
, $\xi(J^\pi_f)$, for the $^{12}{\rm C}(\nu_e,e^-)^{12}{\rm N}$ process
with the conventional neutrino source for the decay-at-rest of muons (DAR)  and
$^{12}{\rm C}({\bar \nu}_e,e^+)^{12}{\rm B}$ reactions with the anti-neutrino
fluxes DAR and those produced by boosted $^6$He ions with $\gamma=6$, $\gamma=10$
and $\gamma=14$.
}%
\begin{ruledtabular}
\newcommand{\cc}[1]{\multicolumn{1}{c}{#1}}
\begin{tabular}{ c | c c | c c c c c }
&\multicolumn{2}{c|}{$(\nu_e,e^-)$}&&&$({\bar \nu}_e,e^+)$&\\
$\xi(J^\pi_f)$&P(I)&P(II)&P(II)&P(I)&&P(I)&\\
&\multicolumn{2}{c|}{DAR}&\multicolumn{2}{c}{DAR}&$\gamma=6$&$\gamma=10$&$\gamma=14$\\ \hline
Allowed&82.6&83.0&79.9&79.9&89.6&77.3&63.9\\
$1^+_1$&45.9&43.4&35.9&36.4&59.5&34.8&23.8
\\
$0^+ $&8.9&   7.3&11.2&13.1&8.6&13.0&13.4
\\
$1^+ $&73.7&75.7 &68.7&66.8&81.0&64.2&50.5
\\
First forbidden&16.9&16.6&19.6&19.6&9.3&21.8&32.9\\
         $0^- $& 0.3& 0.4& 0.7& 0.6&0.4&0.7&0.7
\\
         $1^- $& 8.9& 8.5&11.9&12.3&5.5&13.9&22.0
\\
         $2^- $& 7.7& 7.7&7.0& 6.6&3.4&7.2&10.2
\\
Second forbidden&0.4&0.4&0.5&0.5&1.1&0.9&2.9\\
$2^+ $&0.3&0.3&0.3&0.3&0.8&0.6&1.8
\\
$3^+ $&0.1&0.1&0.2&0.2&0.3&0.3&1.1
\\
Third forbidden&0.0&0.0&0.02&0.02&0.0&0.03&0.2\\
Forth forbidden&0.0&0.0&0.0&0.0&0.0&4$\x 10^{-4}$&0.01\\
\end{tabular}
\end{ruledtabular}
\end{table*}

\begin{figure}[t]
\begin{center}
\vspace{-0.5cm}
{\includegraphics[width=7.cm,height=8.cm]{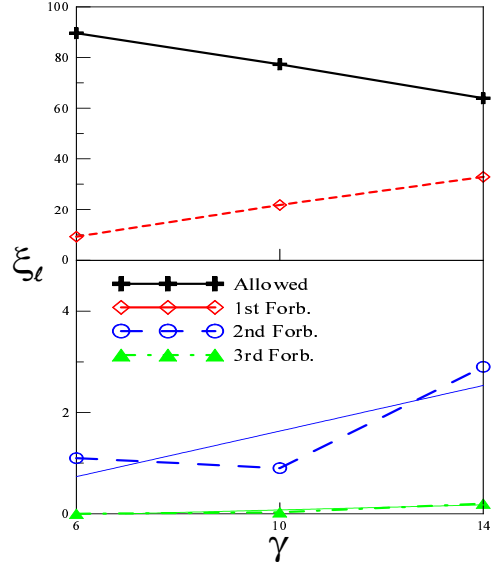}}
\end{center}
\vspace{-1cm}
\caption{\label{fig6} Partial fraction for calculated flux-averaged
antineutrino cross section in $^{12}$C
as function of the $\gamma$-boost. See text for details.
}%
\end{figure}

To analyze possible observables for the excited multipoles we
calculate the folded cross section with the antineutrino fluxes
of $\beta$-beams in a similar way as was discussed in Ref.~\cite{Laz07}.
Table~\ref{tab:2} shows the PQRPA (I) and (II)
calculated flux-averaged cross section with the partial fraction,
\be
\xi_\ell(J^\pi_f)=\left[\frac{\overline{\s}_\ell (J^\pi_f)}
{\overline{\s}^{\rm inc}_\ell}\right] \x 100 \%,
\ee\label{10}
\hspace{-0.32cm}
for the $\nu_e-^{12}$C and ${\bar \nu}_e-^{12}$C
processes with the conventional electron neutrino and antineutrinos
fluxes produced by muon decay-at-rest DAR~\cite{Arm02}
and those produced by boosted $^6$He ions with $\gamma=6$, $\gamma=10$
and $\gamma=14$ adjusting the fluxes of Fig. 4 in Ref.~\cite{Laz07}
to polynomial forms.
From the second and third columns of Table~\ref{tab:2} we note
that $\xi_e(J^\pi_f)$ for $(\nu_e,e^-)$ reactions with the
electron neutrino DAR flux
are quasi-identical, \ie the PQRPA results bring the same
fractional contributions for allowed and forbidden transitions
(the $\xi_e(0^+)$ from P(II) are lower than that for P(I) because
this multipole is most sensitive to the parameters of the $ph$ channel).
A very similar behavior is shown by
$\xi_{\overline e}(J^\pi_f)$ for $(\nu_e,e^+)$ reactions with the
electron antineutrino DAR flux. But in this case the allowed transitions
displace a small part of their contribution to the forbidden transitions
and the third forbidden appears with $\xi_{\overline e}$ non-zero values.
The $\xi_{\overline e}(J^\pi_f)$ for $(\nu_e,e^+)$ reactions with the
electron antineutrino fluxes from $\beta$-beam with
increasing the boost $\gamma$  are shown in the last
three columns of Table~\ref{tab:2}.
According with the increase of $\gamma$, the contribution of allowed
transitions decrease gradually in favor of the first forbidden
transition. In particular this is due to the ground state contribution,
that brings in average half of contribution to
the allowed transition for the different boosts $\gamma$.
In Fig.~\ref{fig6} we show the evolution of the
partial fraction $\xi_{\overline e}$ for allowed,
first forbidden (1st forb.), second forbidden (2nd forb.),
and third forbidden (3rd forb.), flux-averaged
antineutrino cross section in $^{12}$C as a function of the
$\gamma$-boost. The decreasing slope in the allowed
transitions changes to an increasing slope of the
forbidden transition, amounting to $92~\%$, $7~\%$
and $1~\%$ for the 1st forb., 2nd forb. and 3rd forb.
transitions, respectively.

\bigskip

\section{Summarizing conclusions}

The $\nu_e-^{12}$C cross sections calculated in the PQRPA model
are compared with those evaluated in similar RPA-like models.
The exclusive, $\s^{\rm exc}_{\overline e} (E_\nu)$, and
inclusive, $\s^{\rm inc}_{\overline e} (E_\nu)$, antineutrino
cross sections are evaluated for the first time in the PQRPA model.
Theoretical uncertainties of  $\approx 20~ \%$ are seem for
the cross sections with the weak observables in
the $\{^{12}{\rm B},^{12}{\rm C},^{12}{\rm N}\}$ triad.
Ref.~\cite{Kol03} states that it does not matter which nuclear model
is used to evaluate  $\s^{\rm exc}_e (E_\nu)$, as long as
the constrains such as the positron
decay of $^{12}$N, the $\beta$-decay of $^{12}$B, the M1 strength of the
15.11 MeV state in $^{12}$C, and the partial muon capture rate leading
the ground state of $^{12}$B are obeyed. The major of these constrains
(we do not evaluate the M1 strength) and additional data on other
partial muon capture rates are taken into
account in the PQRPA as it is shown in Table~\ref{tab:1}. Nevertheless,
the behavior of exclusive cross section for electron antineutrino
as a function of energy is not so similar
as in the neutrino case. The EFT model seems to move away from the
other models. This could be due to the need to describe
several bound states in  $^{12}$B, whereas in $^{12}$N it
is only the ground state that matters, and the EFT description
avoids the use of the nuclear wave functions, that the other
models employ. This is an important issue to be clarified
because future experiments are using EFT models to estimate
events on supernovae neutrino~\cite{Cad02,Und06,Aga07}.

An explicit comparison of the contribution of the
different multipolarities to  $\s^{\rm inc}_e (E_\nu)$
in the PQRPA is shown with recent results in the RQRPA~\cite{Paa07}.
The contribution of these multipoles to the inclusive
cross section are shown to be similar in both models,
the main contribution comes from the allowed and first
forbidden, and to a smaller amount from the second
forbidden, \ie the contribution from the higher forbiddenness
decreases in a gradual manner. This characteristic behavior is also
displayed by the antineutrino cross sections.

The different behavior of the $\sigma_{e,\mu}^{\rm inc}(E_\nu)$
from RPA, PQRPA, RQRPA and SM with the CRPA and LFG+RPA (it is
also shown for the $\sigma_{\overline e}^{\rm inc}(E_\nu)$)
in the MeV to GeV neutrino energy range
claims detailed experimental and theoretical studies~\cite{Ser04}.
As the main effects are present in the forbidden transitions, we present
the partial fraction in percentage to the inclusive flux-averaged
antineutrino cross section with the antineutrino
fluxes of $\beta$-beams. In the ${\bar \nu}_e-^{12}$C reaction
we note that according with the increase of $\gamma$-boosts, the contribution
of allowed transitions decreases gradually in favor of the
first forbidden transitions. These results enhances
the feasibility of $\beta$-beams to study nuclear
response in low-energy neutrino region.

This work was partially supported by the U.S. DOE grants
DE-FG02-08ER41533 and DE-FC02-07ER41457 (UNEDF, SciDAC-2).

\end{document}